\documentclass[twocolumn,preprintnumbers, amssymb,amsmath,aps,floatfix,prd,nofootinbib,superscriptaddress,showpacs]{revtex4-1}

\usepackage{hyperref}
\usepackage[all]{hypcap}

\usepackage{graphicx}
\usepackage{epstopdf}
\usepackage{amsmath,amsthm,amssymb}

\usepackage{color}

\begin{document}

\title{Extracting jet transport coefficient via single hadron and dihadron productions in high-energy heavy-ion collisions}

\author{Man Xie}
\affiliation{Key Laboratory of Quark and Lepton Physics (MOE) and Institute
of Particle Physics, Central China Normal University, Wuhan 430079, China}

\author{Shu-Yi Wei}
\affiliation{Centre de Physique Th\'eorique, Ecole Polytechnique, CNRS, Universit\'e Paris-Saclay, Route de Saclay, 91128 Palaiseau, France}
\affiliation{Key Laboratory of Quark and Lepton Physics (MOE) and Institute of Particle Physics, Central China Normal University, Wuhan 430079, China}

\author{Guang-You Qin}
\affiliation{Key Laboratory of Quark and Lepton Physics (MOE) and Institute
of Particle Physics, Central China Normal University, Wuhan 430079, China}

\author{Han-Zhong Zhang}
\affiliation{Key Laboratory of Quark and Lepton Physics (MOE) and Institute
of Particle Physics, Central China Normal University, Wuhan 430079, China}

\begin{abstract}

We study the suppressions of high transverse momentum single hadron and dihadron productions in high-energy heavy-ion collisions based on the framework of a next-to-leading-order perturbative QCD parton model combined with the higher-twist energy loss formalism.
Our model can provide a consistant description for the nuclear modification factors of single hadron and dihadron productions in central and non-central nucleus-nucleus collisions at RHIC and the LHC energies.
We quantitatively extract the value of jet quenching parameter $\hat q$ via a global $\chi^2$ analysis, and obtain ${\hat{q}}/{T^3} = 4.1 \sim 4.4$ at $T = 378$~MeV at RHIC and ${\hat{q}}/{T^3} = 2.6 \sim 3.3$ at $T = 486$~MeV at the LHC, which are consistent with the results from JET Collaboration.
We also provide the predictions for the nuclear modification factors of dihadron productions in Pb+Pb collisions at $\sqrt{s_{\rm{NN}}}$ = 5.02 TeV and in Xe+Xe collisions at $\sqrt{s_{\rm{NN}}}$ = 5.44 TeV.

\end{abstract}

\maketitle

\section{Introduction}

The strongly-interacting quark-gluon plasma (QGP) can be created in high-energy heavy-ion collisions performed at the Large Hadron Collider (LHC) and the Relativistic Heavy-Ion Collider (RHIC).
Jet quenching \cite{Gyulassy:1990ye,Wang:1991xy,Qin:2015srf} has been regarded as an extremely useful tool for studying the properties of such hot and dense nuclear matter.
When hard quarks or gluons traverse the QGP matter, they interact with the medium via multiple scatterings and medium-induced gluon radiations.
The elastic and inelastic interactions between jet and medium may cause the energy loss of hard jet and also change the energy distribution among jet partons.
As one of the consequences of jet quenching and energy loss, the yield of high transverse momentum hadrons fragmented from the surviving hard partons is suppressed as compared to that in proton-proton collisions normalized by the number of binary nucleon-nucleon collisions.
Phenomenological studies have been performed on various jet quenching observables, such as the nuclear modifications of single hadron productions \cite{Bass:2008rv, Armesto:2009zi, Burke:2013yra, Cao:2017hhk}, dihadron and photon-hadron correlations \cite{Zhang:2007ja, Majumder:2004pt, Qin:2009bk, Renk:2008xq, Chen:2016vem, Chen:2017zte}, as well as the observables related to fully reconstructed jets in relativistic nuclear collisions \cite{Qin:2010mn,CasalderreySolana:2010eh,He:2011pd,Young:2011qx,Zapp:2012ak,Wang:2013cia,Chang:2016gjp,Tachibana:2017syd}.

In recent years, jet quenching studies have entered the quantitative era in that much effort has been devoted to the quantitative extraction of the so-called jet quenching parameter $\hat{q}$.
This parameter is defined as the transverse momentum squared per unit length exchanged between the propagating hard parton and the traversed medium, $\hat{q} = d\langle (\Delta p_{T})^2 \rangle /dt$, and may be directly related to the gluon density of the nuclear medium \cite{Baier:1996sk}.
Jet transport parameter $\hat{q}$ also controls the amount of medium-induced gluon radiation and thus radiative jet energy loss \cite{Baier:1996kr, Baier:1996sk, Baier:1998kq, Guo:2000nz, Wang:2001ifa, Majumder:2009ge}.
In addition, the transverse momentum broadening effect as controlled by $\hat{q}$ may lead to significant nuclear modification on back-to-back dijet, dihadron and other jet-related angular correlations \cite{Chen:2016vem,Chen:2017zte}.
Among many quantitative jet quenching studies, one of the most important steps is performed by JET Collaboration in Ref. \cite{Burke:2013yra} which has compared five different theoretical jet quenching models with the nuclear modification data on single hadron productions in most central collisions at RHIC and the LHC and quantitatively extracted the temperature dependence of jet quenching parameter $\hat{q}$.
The values of $\hat{q}$ temperatures available at RHIC and the LHC have been obtained as: $\hat{q}/T^3 = 4.6\pm 1.2$ at $T\approx 370$~MeV and $\hat{q}/T^3 = 3.7\pm 1.4$ at $T \approx 470$~MeV for a $10$~GeV quark jet \cite{Andres:2016iys}.
Following this direction, Ref \cite{Andres:2016uio,  Andres:2017awo} has studied the centrality and collision energy dependence of $\hat{q}$ values at both RHIC and the LHC.  Also, Refs. \cite{Chen:2016vem} has utilized the nuclear modification data on back-to-back dihadron and hadron-jet angular correlations to extract the value of $\hat{q}$ at RHIC.

This paper follows closely the above efforts and study the nuclear modifications of both single hadron and dihadron productions at high transverse momenta using a next-to-leading-order (NLO) perturbative QCD model combined with the higher-twist energy loss formalism.
In particular, we perform a global $\chi^2$ analysis on the nuclear modification data on single hadron and dihadron productions at RHIC \cite{Adare:2008qa,Adare:2012wg,Adams:2006yt,STAR:2016jdz} and the LHC \cite{Abelev:2012hxa,CMS:2012aa,Khachatryan:2016odn,Acharya:2018qsh,Acharya:2018eaq,Adam:2016xbp,Aamodt:2011vg,Conway:2013xaa} and quantitatively extract the values of jet quenching parameter $\hat q$.
Our analysis yields ${\hat{q}}/{T^3} = 4.1 \sim 4.4$ at $T = 378$~MeV at RHIC and ${\hat{q}}/{T^3} = 2.6 \sim 3.3$ at $T = 486$~MeV at the LHC.
These results are quantitatively consistent with JET Collaboration.
We also extract the $\hat{q}$ values for Pb+Pb collisions at $\sqrt{s_{\rm{NN}}} = 5.02$~TeV and Xe+Xe collisions at $\sqrt{s_{\rm{NN}}} = 5.44$~TeV using the single hadron nuclear modification data, and predict the nuclear modification factors for dihadron productions for these collisions.

Our paper is organized as follows. In Sec. II, we briefly introduce our framework to study the productions of single hadrons and dihadrons at high transverse momenta in proton-proton and nucleus-nucleus collisions. In Sec. III, we perform a global $\chi^2$ analysis and extract jet quenching parameter $\hat{q}$ from the nuclear modification data on single hadron and dihadron productions at RHIC and the LHC. We also provide our predictions for the nuclear modification factors of dihadron productions in central and non-central Pb+Pb collisions at $\sqrt{s_{\rm{NN}}} = 5.02$~TeV and Xe+Xe collisions at $\sqrt{s_{\rm{NN}}} = 5.44$~TeV at the LHC.
Sec. IV contains our summary.

\section{Framework}

In high-energy proton-proton collisions, the production cross section of high transverse momentum hadrons can be factorized into a convolution of parton distribution functions (PDFs), the cross section of hard partonic scatterings, and fragmentation functions (FFs),
\begin{eqnarray}
	\frac{d\sigma_{pp}^h}{dyd^2p_T}&=&\sum_{abcd}\int dx_a dx_b f_{a/p}(x_a,\mu^2) f_{b/p}(x_b,\mu^2)\nonumber \\
&&\times
	\frac{1}{\pi}\frac{d\sigma_{ab\rightarrow cd}}{d\hat{t}}\frac{D_{c}^{h}(z_c,\mu^2)}{z_c}+\mathcal {O}(\alpha_s^3).
\label{eq:pp-sin-spec}
\end{eqnarray}
Here, $f_a(x_a,\mu^2)$ and $f_b(x_b,\mu^2)$ are parton distribution functions which we take from CT14 \cite{Hou:2016nqm}; $D_{c}^{h}(z_c,\mu^2)$ is fragmentation function which we take from Refs.~\cite{Kretzer:2000yf,Wang:2004yv}; $d\sigma_{ab\rightarrow cd}/d\hat{t}$ is the tree-level $2 \to 2$ partonic scattering cross section. The NLO correction at $\mathcal {O}(\alpha_s^3)$ contains $2\rightarrow2$ virtual diagrams and $2\rightarrow3$ tree diagrams, and has been included in our calculation. It has been shown in Ref.~\cite{Zhang:2007ja} that NLO perturbative QCD calculation for single $\pi^0$ production in proton-proton collisions agrees well with the experimental data at RHIC.

Similarly, the production cross section for high transverse momentum dihadrons in high-energy proton-proton collisions can be written as,
\begin{eqnarray}
\frac{d\sigma_{pp}^{h_1h_2}}{dPS}
	&=& \sum_{abcd}\int \frac{dz_c}{z_c^2} \frac{dz_d}{z_d^2} x_a f_{a/p}(x_a, \mu^2) x_b f_{b/p}(x_b, \mu^2) \nonumber\\
	&&\times \frac{1}{\pi} \frac{d\sigma_{ab\rightarrow cd}}{d\hat{t}}
	D_{c}^{h_1}(z_c, \mu^2) D_{d}^{h_2}(z_d, \mu^2)
	\nonumber\\
	&&\times \delta^2 (\frac{\vec{p}_T^{h_1}}{z_c} + \frac{\vec{p}_T^{h_2}}{z_d}) + \mathcal {O}(\alpha_s^3),
\end{eqnarray}
where the phase space is $dPS = dy^{h_1} d^2 p_T^{h_1} dy^{h_2} d^2 p_T^{h_2}$.

In relativistic nucleus-nucleus collisions, one has to consider both cold nuclear matter effect in the initial state and hot nuclear matter effect in the final state.
The yield of single hadron production at high transverse momentum may be obtained as \cite{Zhang:2007ja,Zhang:2009rn},
\begin{eqnarray}
\frac{dN_{AB}^h}{dyd^2p_T}&&=\sum_{abcd}\int dx_adx_b d^2r
t_A(\vec{r}) t_B(\vec{r}+\vec{b}) \nonumber \\
&&\times\
f_{a/A}(x_a,\mu^2,\vec{r}) f_{b/B}(x_b,\mu^2,\vec{r}+\vec{b}) \nonumber \\
&&\times \frac{1}{\pi}
\frac{d\sigma_{ab\rightarrow cd}}{d\hat{t}}
	\frac{\tilde{D}_{c}^{h}(z_{c},\mu^2,\Delta E_c)}{z_c} +\mathcal {O}(\alpha_s^3). \ \
\label{eq:AA-sin-spec}
\end{eqnarray}
Similarly, the yield of dihadron production at high transverse momentum in nucleus-nucleus collisions may be calculated as  \cite{Zhang:2007ja,Zhang:2009rn,Wang:2003mm}
\begin{eqnarray}
\frac{dN_{AB}^{h_1h_2}}{dPS} &&=
\sum_{abcd}\int{\frac{dz_c}{z^2_c} \frac{dz_d}{z^2_d}}
 d^2r t_A(\vec{r}) t_B(\vec{r}+\vec{b}) \nonumber\\
&&\times x_a f_{a/A}(x_a,\mu^2,\vec{r}) x_b f_{b/B}(x_b,\mu^2,\vec{r}+\vec{b}) \nonumber\\
	&&\times \frac{1}{\pi}{\frac{d\sigma_{ab{\rightarrow}cd} }{d\hat{t}}}
	\tilde{D}_{c}^{h_1} (z_c,\mu^2,\Delta E_c) \tilde{D}_{d}^{h_2}(z_d,\mu^2,\Delta E_d) \nonumber\\
&&\times \delta^2 (\frac{\vec p_T^{h_1}}{z_c} + \frac{\vec{p}_T^{h_2}}{z_d}) +O({\alpha}_s^3).
\label{eq:AA-di-spec}
\end{eqnarray}
In the above two equations, $t_A(\vec{r})$ is the nuclear thickness function, normalized as $\int d^2r t_A(\vec{r}) = A$, with $A$ the mass number of the nucleus.
Here we use the Woods-Saxon form for the nuclear density distribution.
$f_{a/A}(x_a,\mu^2,\vec{r})$ is the nuclear modified PDF, which we calculate as follows \cite{Wang:1996yf,Li:2001xa}:
\begin{eqnarray}
f_{a/A}(x_a,\mu^2,\vec{r}) &&= S_{a/A}(x_a,\mu^2,\vec{r})\left[\frac{Z}{A}f_{a/p}(x_a,\mu^2)\right. \nonumber\\
&&+\left.\left(1-\frac{Z}{A}\right)f_{a/n}(x_a,\mu^2)\right],
\end{eqnarray}
where $Z$ is the proton number of the nucleus.
Here, $S_{a/A}(x_a,\mu^2,\vec{r})$ is called the nuclear shadowing factor and denotes the nuclear modification to the PDF in a free proton $f_{a/p}(x_a,\mu^2)$. The shadowing factor $S_{a/A}(x_a,\mu^2,\vec{r})$ is calculated using the following form \citep{Emelyanov:1999pkc,Hirano:2003pw},
\begin{eqnarray}
S_{a/A}(x_a,\mu^2,\vec{r})=1+[S_{a/A}(x_a,\mu^2)-1] \frac{At_A(\vec{r})}{\int{d^2}r [t_A(\vec{r})]^2}, \ \ \
\end{eqnarray}
where $S_{a/A}(x_a,\mu^2)$ is taken from the EPPS16 \cite{Eskola:2016oht}.
$\tilde{D}_{c}^{h}(z_{c},\Delta E_c)$ is the medium-modified fragmentation function and is calcualted as follows \cite{Wang:2004yv,Zhang:2007ja,Zhang:2009rn}:
\begin{eqnarray}
	&& \tilde{D}_{c}^{h}(z_c,\mu^2,\Delta{E_c}) = (1-e^{-\langle{N_g}\rangle})\left[\frac{z'_c}{z_c}D_{h}^{c}(z'_c,\mu^2)\right. \nonumber\\
&& \phantom{XX}+\left.{\langle{N_g}\rangle}\frac{{z_g}'}{z_c}D_{g}^{h}({z_g}',\mu^2)\right]+e^{-\langle{N_g}\rangle}D_{c}^{h}({z_c},\mu^2),
\end{eqnarray}
where $\Delta E_c$ is the energy loss of parton $c$, $z_c = p_T/p_{Tc}$, $z'_c = p_{T}/(p_{Tc} - \Delta E_c)$, $z'_g =\langle{N_g}\rangle p_T/\Delta{E_c}$ and $\langle N_g \rangle$ is the average number of gluons radiated by parton $c$.
In this work, we use the higher twist formalism \cite{Wang:2009qb,Wang:2001cs,Wang:2002ri} to calculate medium-induced gluon radiation and parton energy loss.
For a quark with initial energy $E$, the total energy loss $\Delta E$ can be calculated as,
\begin{eqnarray}
\frac{\Delta{E}}{E} &=& \frac{2C_A\alpha_s}{\pi} \int d\tau \int \frac{dl_T^2}{l_T^4}\int dz \nonumber\\
	&&\times \left[1+(1-z)^2\right] \hat{q} \sin^2(\frac{l_T^2\tau}{4z(1-z)E}),
\end{eqnarray}
where $C_A=3$, and $l_T$ is the transverse momentum of radiated gluon. We assume the energy loss of a gluon is simply $9/4$ times that of a quark \cite{Wang:2009qb}.
The average number of radiated gluons from the propagating hard parton is calculated as \cite{Chang:2014fba},
\begin{eqnarray}
	\langle N_g \rangle &=& \frac{2C_A \alpha_{s}}{\pi} \int d\tau \int \frac{dl_T^2}{l_T^4}\int \frac{dz}{z} \nonumber\\
	&&\times \left[1+(1-z)^2\right] \hat{q} \sin^2(\frac{l_T^2\tau}{4z(1-z)E}).
\end{eqnarray}
The parton energy loss is controlled by jet transport parameter $\hat{q}$ \citep{Baier:1996sk}, for which we take the following form:
\begin{eqnarray}
\hat q = \hat q_0 \frac{T^3}{T_0^3}\frac{p^{\mu}u^{\mu}}{p_0},
\end{eqnarray}
where $T$ is the local temperature of the medium, $T_0$ is a reference temperature which is usually taken as the temperature at the center of the medium at the hydrodynamics initial time $\tau_0 = 0.6$~fm in central nucleus-nucleus collisions, and $u^{\mu}$ is the four flow velocity of the fluid. In our calculation, the dynamical evolution of the QGP medium is obtained using the OSU (2+1)-dimensional viscous hydrodynamics model (VISH2+1) \cite{Song:2007fn,Song:2007ux,Qiu:2011hf,Qiu:2012uy}.

\section{Numerical results}

In this section, we present our numerical results for single hadron and dihadron nuclear modification factors in Au+Au collisions at $\sqrt{s_{\rm{NN}}}=0.2$~TeV, Pb+Pb collisions at $\sqrt{s_{\rm{NN}}}=2.76$~TeV and $5.02$~TeV, and Xe+Xe collisions at $\sqrt{s_{\rm{NN}}}=5.44$~TeV.
A global $\chi^2$ analysis is performed to extract the jet quenching parameter $\hat{q}$ in different collision systems and different collision energies at RHIC and the LHC.
Based on our analysis, we also provide the predictions for the nuclear modification factors of dihadron productions in Pb+Pb collisions at $\sqrt{s_{\rm{NN}}} = 5.02$~TeV and Xe+Xe collisions at $\sqrt{s_{\rm{NN}}} = 5.44$~TeV.

The nuclear modification factor $R_{AA}$ for single hadron production in heavy-ion collisions is defined as \cite{Wang:2004yv},
\begin{eqnarray}
R_{AA}=\frac{dN_{AB}^h/dyd^2p_T}{\langle T_{AA}\rangle d{\sigma}_{pp}^h/dyd^2p_T},
\end{eqnarray}
where $T_{AA}(\vec{b}) =\int d^2r t_A(\vec{r})t_B(\vec{r}+\vec{b})$ is the overlap function of two colliding nuclei and the average in the equation is taken for a given centrality class.

As for dihadron production at high transverse momentum in heavy-ion collisions, the nuclear modification factor $I_{AA}$ can be defined either as a function of $p_T^{assoc}$ or as a function of $z_T=p_T^{assoc}/p_T^{trig}$ \cite{Zhang:2007ja}
\begin{eqnarray}
	&& I_{AA}(p_T^{assoc}) = \frac{D_{AA}(p_T^{assoc})}{D_{pp}(p_T^{assoc})},
\nonumber\\
	&& I_{AA}(z_T) = \frac{D_{AA}(z_T)}{D_{pp}(z_T)},
\end{eqnarray}
where $D_{AA}(z_T) = p_T^{trig} D_{AA}(p_T^{assoc})$ is called hadron-triggered fragmentation function \cite{Wang:2003aw},
\begin{eqnarray}
	D_{AA}(z_T)=p_T^{trig}\frac{d{N}_{AA}^{h_1h_2}/dy^{trig}dp_T^{trig}dy^{assoc}dp_T^{assoc}}{\langle T_{AA}\rangle d{\sigma}_{AA}^{h_1}/dy^{trig}dp_T^{trig}}. \ \ \ \
\end{eqnarray}

\begin{figure}[tbh]
\begin{center}
\includegraphics[width=0.45\textwidth]{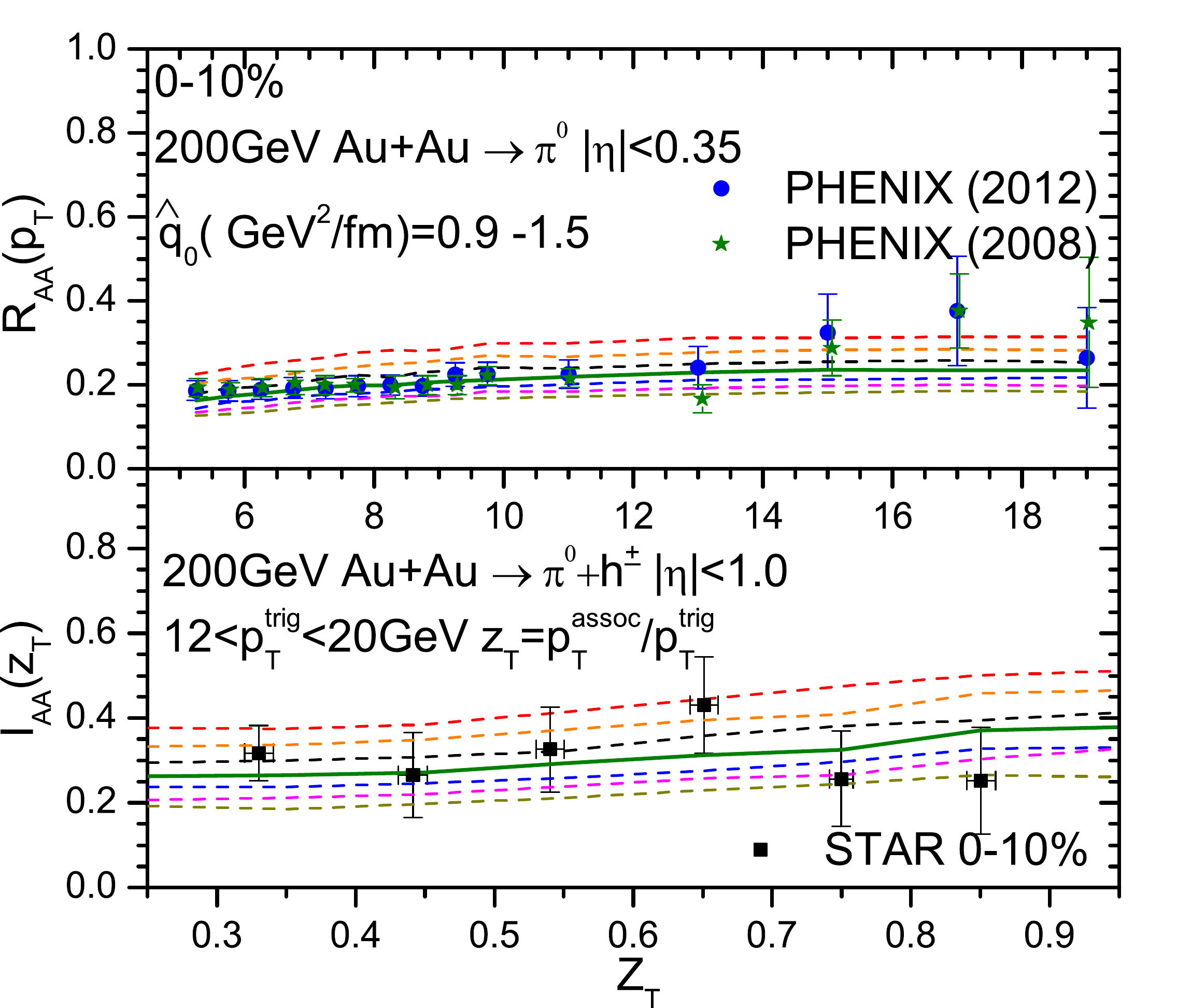}
\end{center}
\caption[*]{The single hadron and dihadron suppression factors in central $0-10\%$ Au+Au collisions at $\sqrt{s_{\rm NN}}=0.2$~TeV compared with PHENIX \cite{Adare:2008qa,Adare:2012wg} and STAR \cite{STAR:2016jdz} data.}
\label{fig:rhic0-10}
\end{figure}

\begin{figure}[tbh]
\begin{center}
\includegraphics[width=0.45\textwidth]{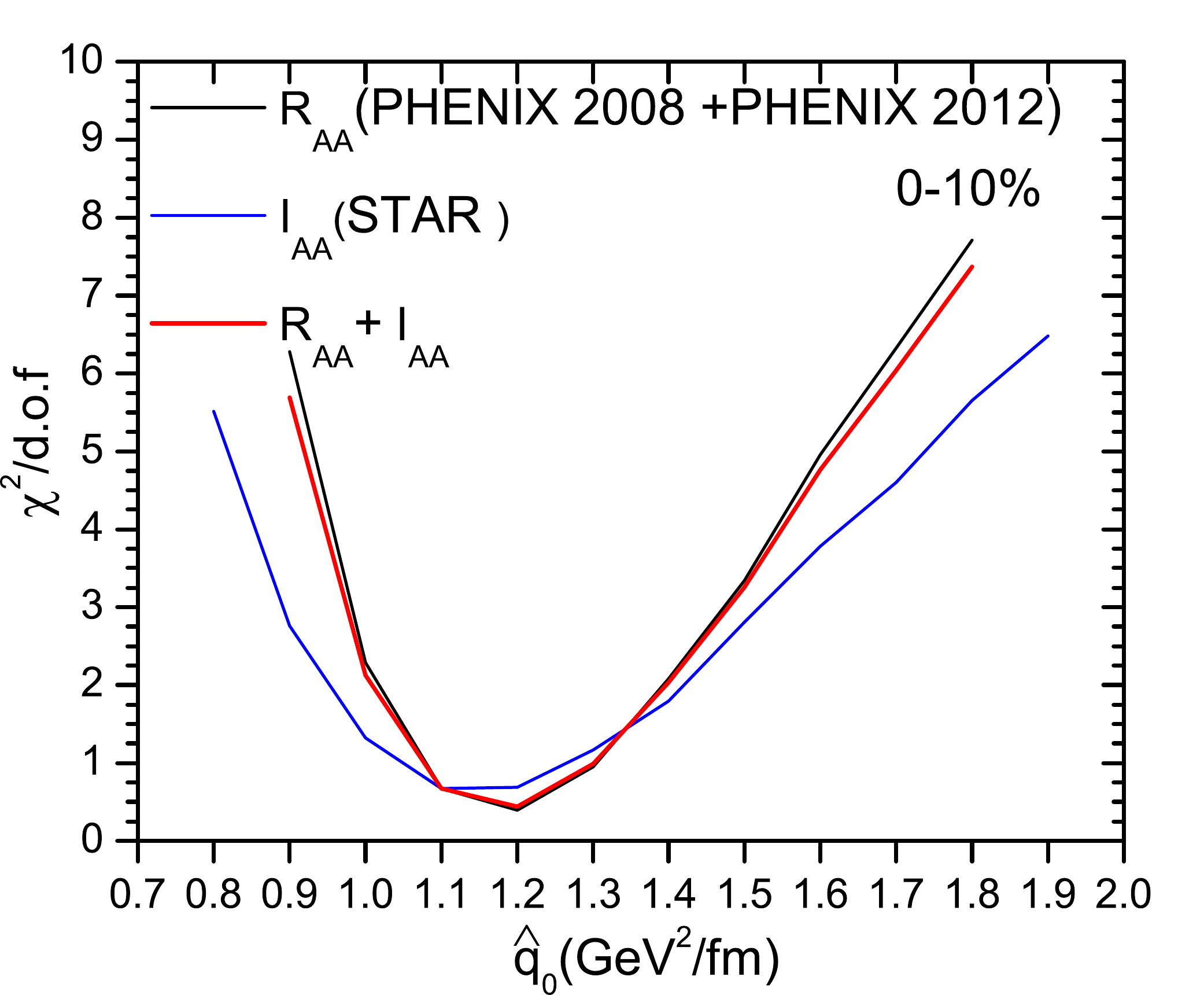}
\end{center}
	\caption[*]{Global $\chi^2$ analysis for single hadron and (or) dihadron nuclear suppression factors in Au+Au collisions at $\sqrt{s_{\rm{NN}}}=0.2$~TeV at RHIC.}
\label{fig:rhic0-10-x2}
\end{figure}


\begin{figure}[tbh]
\begin{center}
\includegraphics[width=0.45\textwidth]{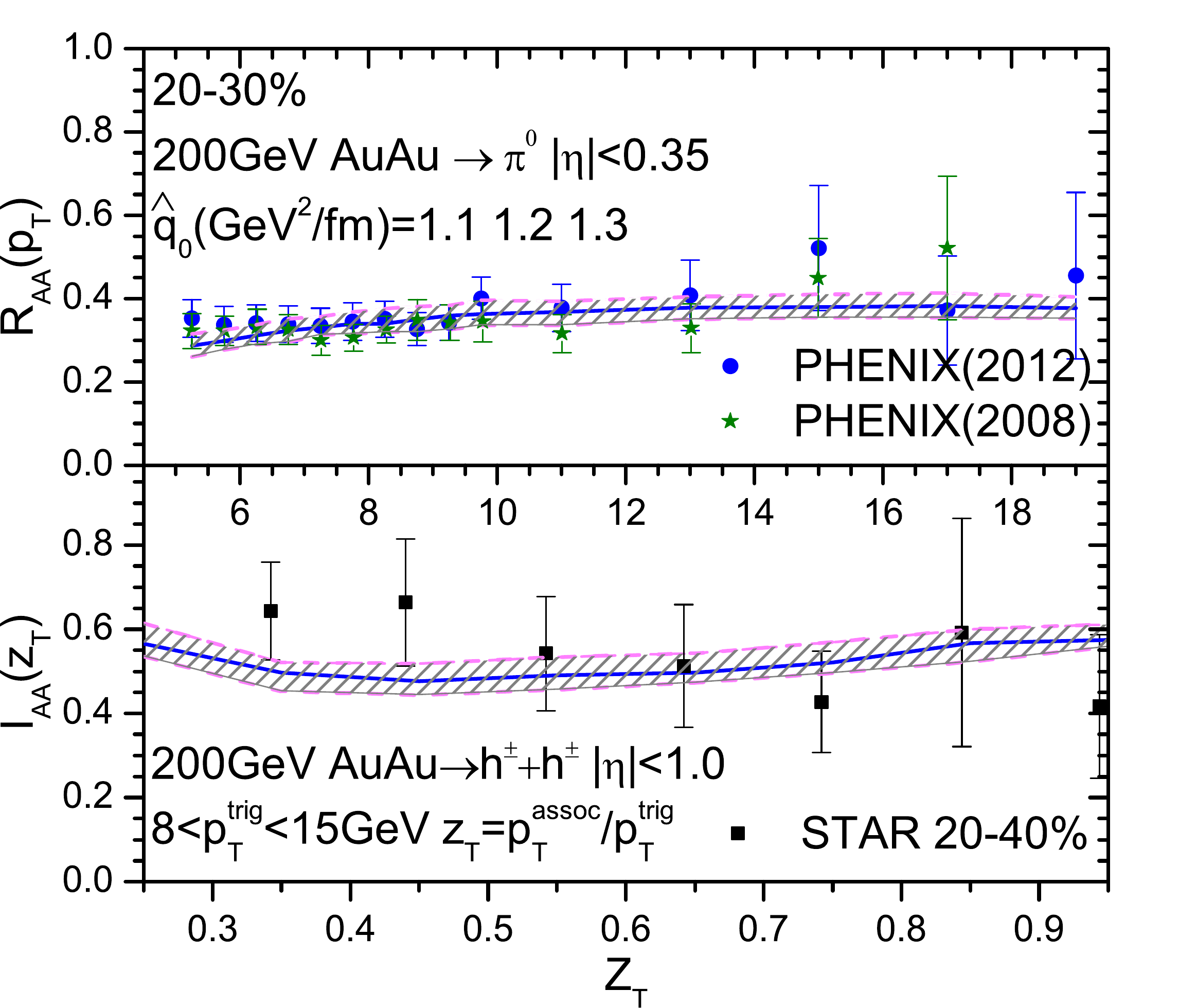}
\end{center}
	\caption[*]{The single hadron and dihadron suppression factors in mid-central Au+Au collisions at $\sqrt{s_{\rm{NN}}}=0.2$~TeV compared with PHENIX \cite{Adare:2008qa,Adare:2012wg} and STAR \cite{Adams:2006yt} data.}
\label{fig:rhic20-30}
\end{figure}

\subsection{Au+Au collisions at $\sqrt{s_{\rm{NN}}}=0.2$~TeV at RHIC}

Fig. \ref{fig:rhic0-10} shows our calculations for single hadron and dihadron nuclear modification factors in central ($0-10\%$) Au+Au collisions at $\sqrt{s_{\rm{NN}}}=0.2$~TeV at RHIC compared with the experimental data taken from PHENIX \cite{Adare:2008qa,Adare:2012wg} and STAR \cite{STAR:2016jdz} Collaborations.
In each plot, different lines represent our model calculations for $R_{AA}$ or $I_{AA}$ using different values of jet quenching parameter $\hat{q}_0$.
The solid line in the middle denotes the result using the best value of $\hat{q}_0$ obtained from our global $\chi^2$ analysis, which is shown in Fig. \ref{fig:rhic0-10-x2}.
In the figure, we also show $\chi^2/{\rm d.o.f}$ as a function of $\hat{q}_0$ using only $R_{AA}$ data or only $I_{AA}$ data.
We can see that two fitting results are consistent with each other.
This means that with the similar value of $\hat{q}_0$, both single hadron and dihadron nuclear suppression factors can be described consistently within our jet energy loss model.
Our global $\chi^2$ analysis renders: $\hat{q}_0=1.1 \sim 1.2$~GeV$^2/$fm at $T_0=378$~MeV. In terms of the scaled dimensionless jet quenching parameter, it reads, $\hat{q}/T^3 = 4.1 \sim 4.4$ at $T=378$~MeV.
These values are consistent with the results obtained by JET Collaboration \cite{Burke:2013yra}.

To test the goodness of our approach, we use the same $\hat{q}_0$ value obtained above to calculate the nuclear modification factors $R_{AA}$ and $I_{AA}$ in mid-central Au+Au collision at $\sqrt{s_{\rm{NN}}}=0.2$~TeV at RHIC.
The result is shown in Fig. \ref{fig:rhic20-30}, where the solid lines in the middle denote the results using the best $\hat{q}_0$ value (i.e, $\hat{q}_0 = 1.2$~GeV$^2$/fm at $T_0=378$~MeV), while the other two lines represent the uncertainty for the extracted $\hat{q}_0$ value ($\hat{q}_0=1.1$ or $1.3$ ~GeV$^2$/fm for the two lines).
We can see that with the similar $\hat{q}_0$ value, our model can provide a good description of experimental data on single and dihadron nuclear modification in both central and non-central Au+Au collisions  at $\sqrt{s_{\rm{NN}}}=0.2$~TeV at RHIC.

\begin{widetext}

\begin{figure}[tbh]
\begin{center}
\includegraphics[width=0.45\textwidth, height=0.56\textwidth]{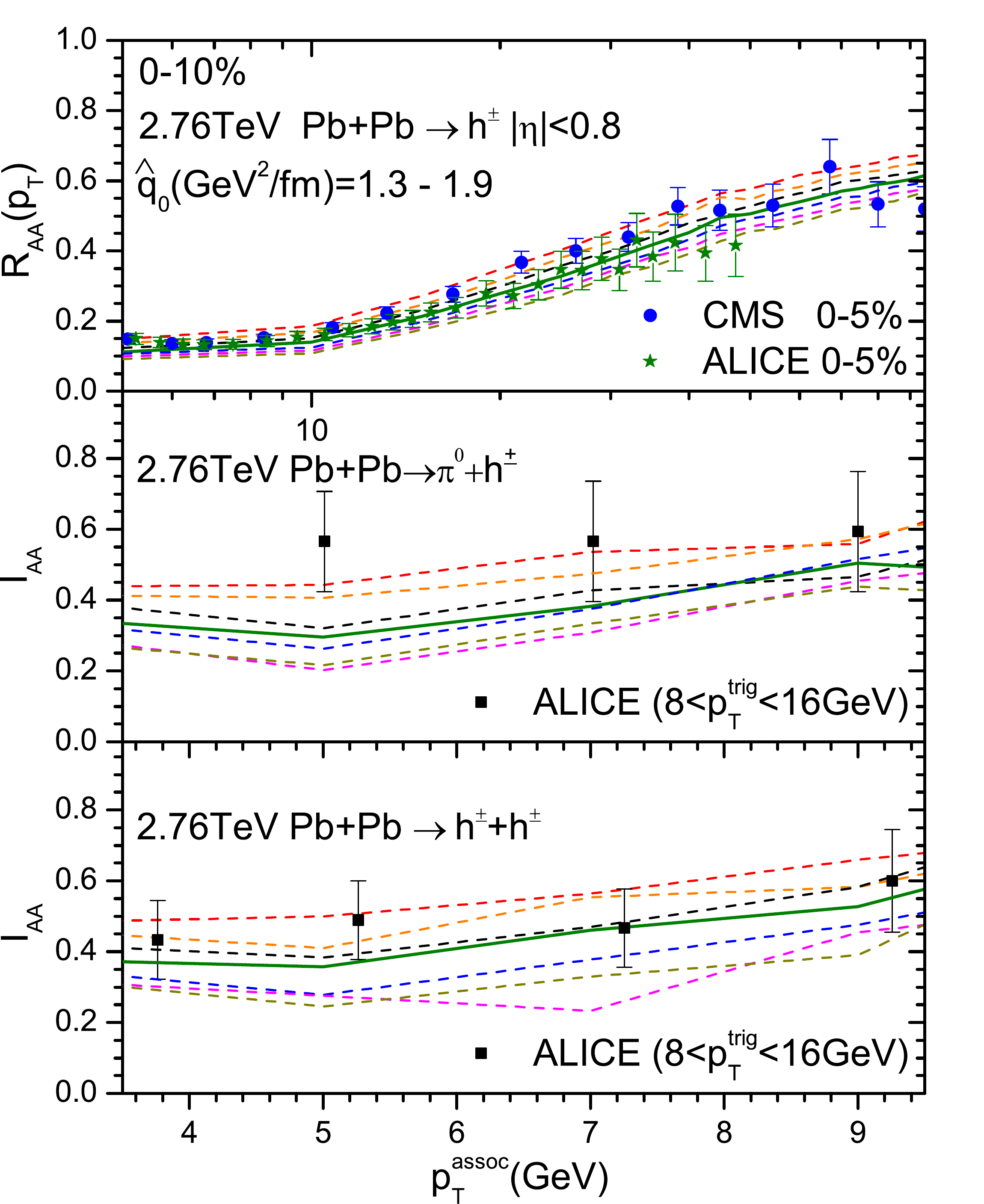}
\includegraphics[width=0.45\textwidth, height=0.54\textwidth]{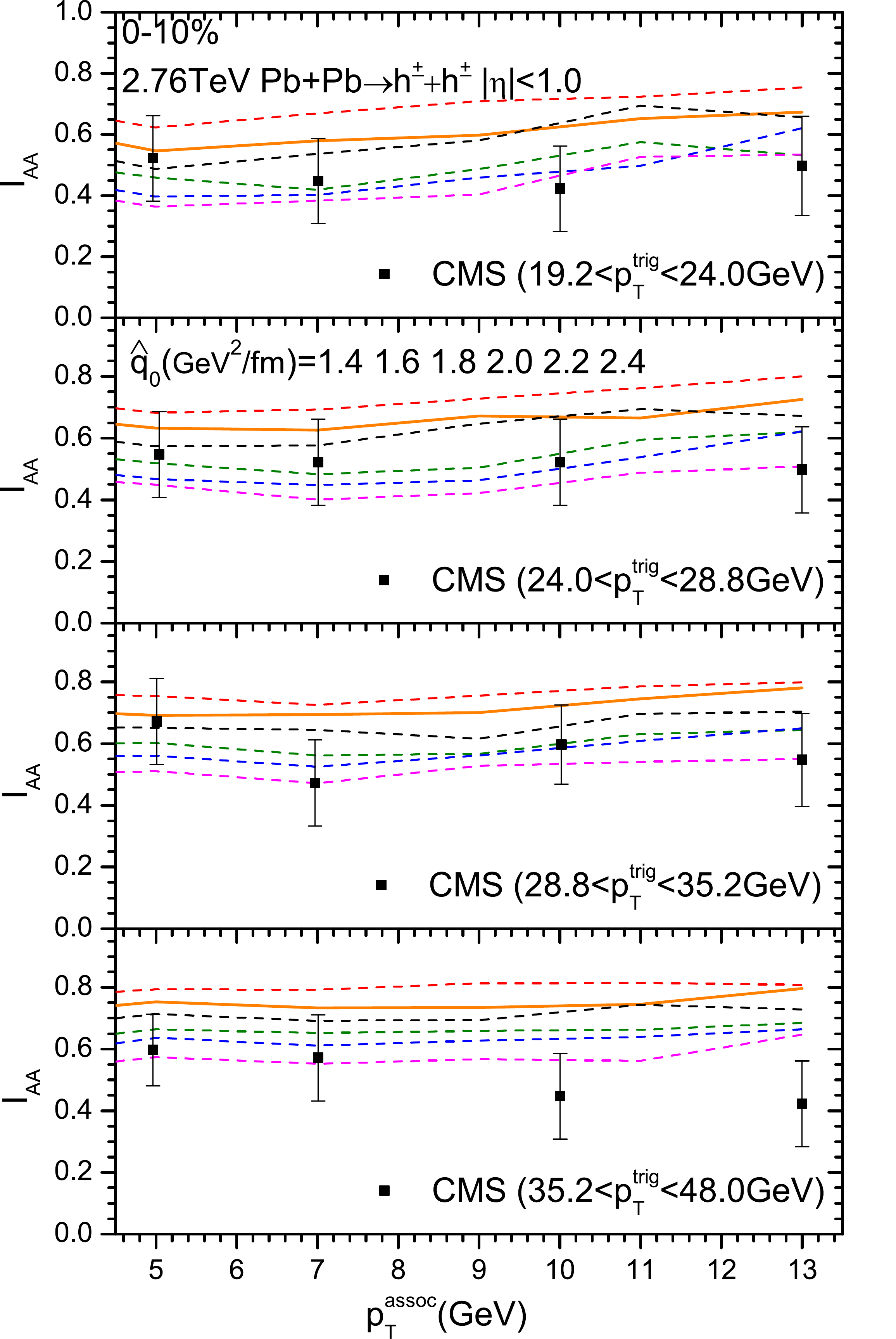}
\end{center}
\caption[*]{The single hadron and dihadron suppression factors in  0-10\% Pb+Pb collisions at $\sqrt{s_{\rm{NN}}}=2.76$ TeV compared with CMS \cite{CMS:2012aa, Conway:2013xaa} and ALICE \cite{Abelev:2012hxa,Aamodt:2011vg,Adam:2016xbp} data.}
\label{fig:lhc276IAA0-10-1}
\end{figure}

\end{widetext}

\subsection{Pb+Pb collisions at $\sqrt{s_{\rm{NN}}}=2.76$~TeV at the LHC}

\begin{figure}[tbh]
\begin{center}
\includegraphics[width=0.45\textwidth]{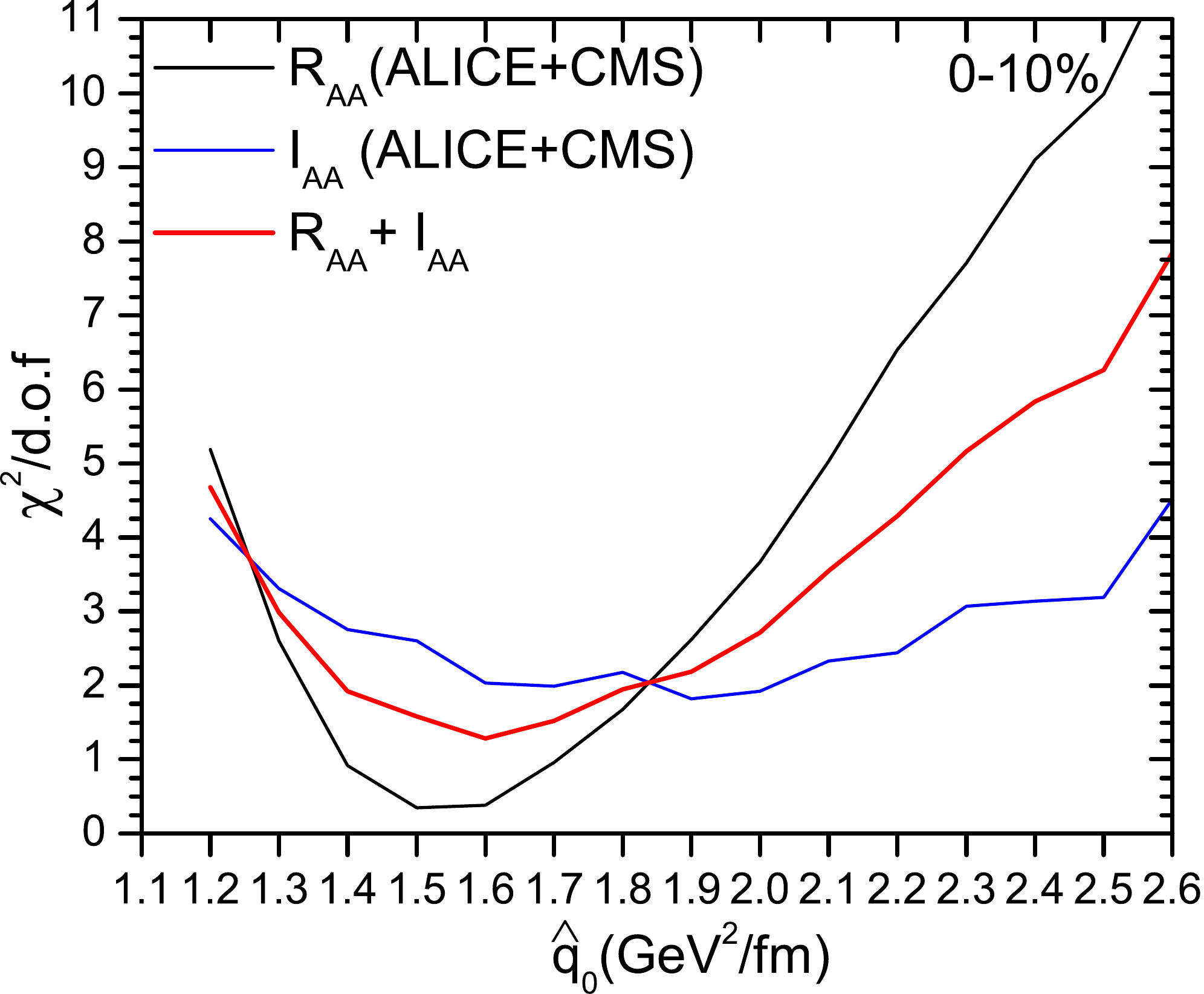}
\end{center}
\caption[*]{Global $\chi^2$ analysis for single hadron and (or) dihadron nuclear suppression factors in Pb+Pb collisions at $\sqrt{s_{\rm{NN}}}=2.76$~TeV at the LHC.}
\label{fig:lhc2760x2}
\end{figure}

Now we present our numerical results for Pb+Pb collisions at $\sqrt{s_{\rm{NN}}}=2.76$~TeV at the LHC.
Fig. \ref{fig:lhc276IAA0-10-1} shows our calculations for single hadron and dihadron nuclear modification factors in central ($0-10\%$) Pb+Pb collisions at $\sqrt{s_{\rm{NN}}}=2.76$~TeV compared with the experimental data from ALICE \cite{Abelev:2012hxa,Aamodt:2011vg,Adam:2016xbp} and CMS \cite{CMS:2012aa,Conway:2013xaa} Collaborations.
In each plot, different lines represent our model calculations for $R_{AA}$ and (or) $I_{AA}$ using different $\hat{q}_0$ values.
The solid line in the middle denotes the results using the best $\hat{q}_0$ value obtained from our global $\chi^2$ analysis.
Also show in Fig. \ref{fig:lhc2760x2} is the $\chi^2$ analysis of $\hat{q}_0$ value using only $R_{AA}$ data or $I_{AA}$ data.
Although there is some small difference between two fitting results, they are quantitatively consistent with each other within the uncertainties.
Such consistency implies that with the similar values of $\hat{q}_0$, our jet energy loss model can provide a consistent description of both single hadron and dihadron nuclear suppression factors in Pb+Pb collisions at  $\sqrt{s_{\rm{NN}}}=2.76$~TeV.
From Fig. \ref{fig:lhc2760x2}, we obtain: $\hat{q}_0 =1.5 \sim 1.9$~GeV$^2/$fm at $T_0=486$~MeV, which translates into the scaled jet quenching parameter, $\hat{q}/T^3 = 2.6 \sim 3.3$ at $T=486$~MeV.
This values are also consistent with JET Collaboration \cite{Burke:2013yra}.

\begin{figure}[tbh]
\begin{center}
\includegraphics[width=0.45\textwidth]{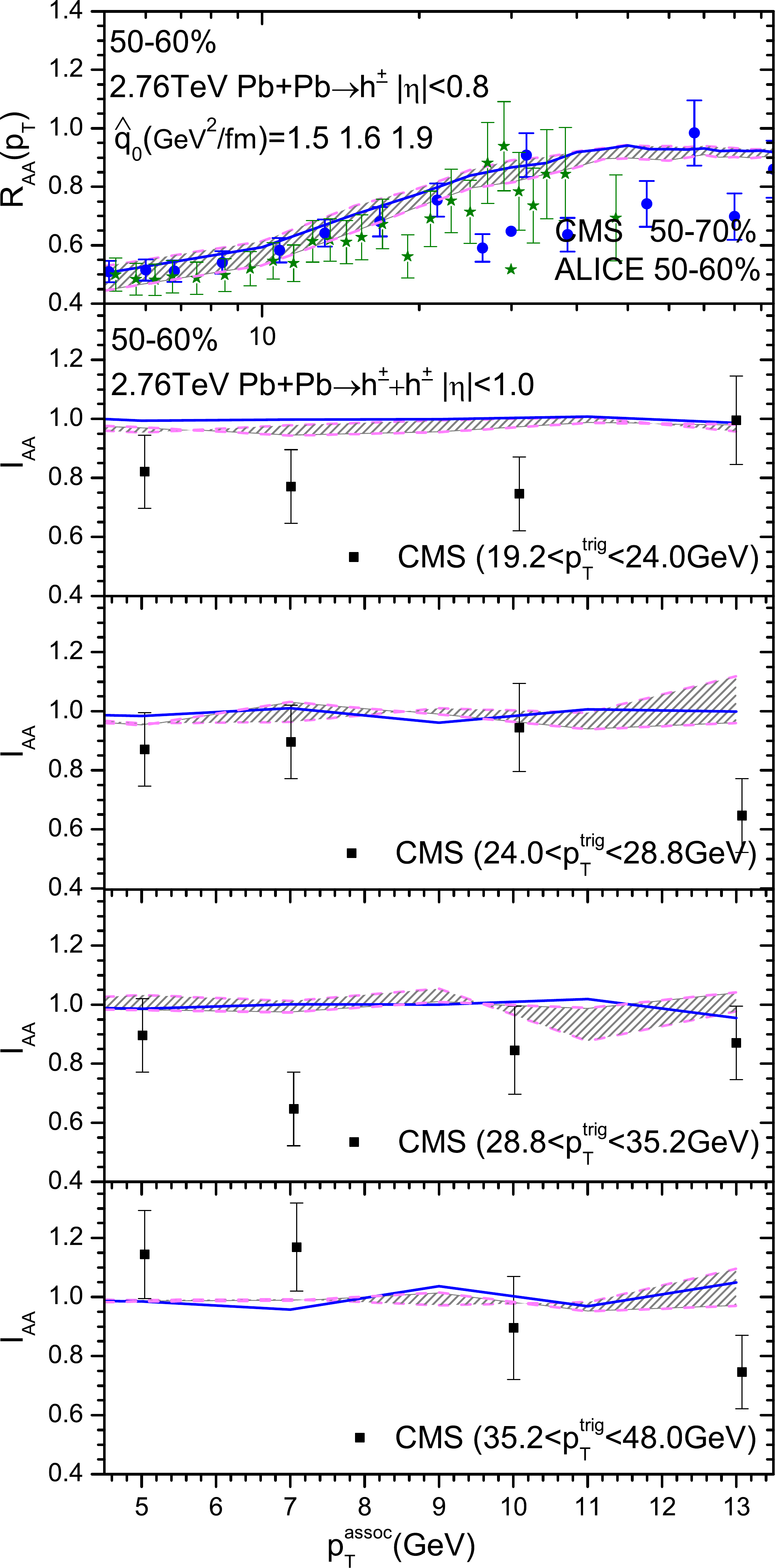}
\end{center}
	\caption[*]{The single hadron and dihadron suppression factors in non-central (50-60\%) Pb+Pb collisions at $\sqrt{s_{\rm{NN}}}=2.76$ TeV compared with ALICE \cite{Abelev:2012hxa} and CMS \cite{CMS:2012aa,Conway:2013xaa} data.}
\label{fig:lhc276IAA50-60}
\end{figure}

We also test our approach by using the same $\hat{q}_0$ value obtained above to calculate the nuclear modification factors $R_{AA}$ and $I_{AA}$ in the non-central ($50-60\%$) Pb+Pb collision at $\sqrt{s_{\rm{NN}}}=2.76$~TeV.
The result is shown in Fig. \ref{fig:lhc276IAA50-60}: the solid lines in the middle denote the results using the best $\hat{q}_0$ value (i.e., $\hat{q}_0 = 1.6$~GeV$^2$/fm at $T_0=486$~MeV), while the other two lines (using $\hat{q}_0 = 1.5$ and $1.9$~GeV$^2$/fm) represent the uncertainty for our extracted $\hat{q}_0$ value.
We can see that with the same $\hat{q}_0$ value, our jet energy loss model can also describe the experimental data on single and dihadron nuclear modification in non-central Pb+Pb collisions at $\sqrt{s_{\rm{NN}}}=2.76$~TeV.
Another interesting result is that for both Au+Au collisions at RHIC and Pb+Pb collision at the LHC, the nuclear modification factors $I_{AA}$ for dihadron productions are typically larger than single hadron suppression factors $R_{AA}$ given the same nucleus-nucleus collision conditions.
One of the main reasons for such difference is the dominance of tangential emissions in dijet (dihadron) events, as has been been pointed out in Ref. \cite{Zhang:2007ja}.

\begin{widetext}

\begin{figure}[tbh]
\begin{center}
\includegraphics[width=0.45\textwidth]{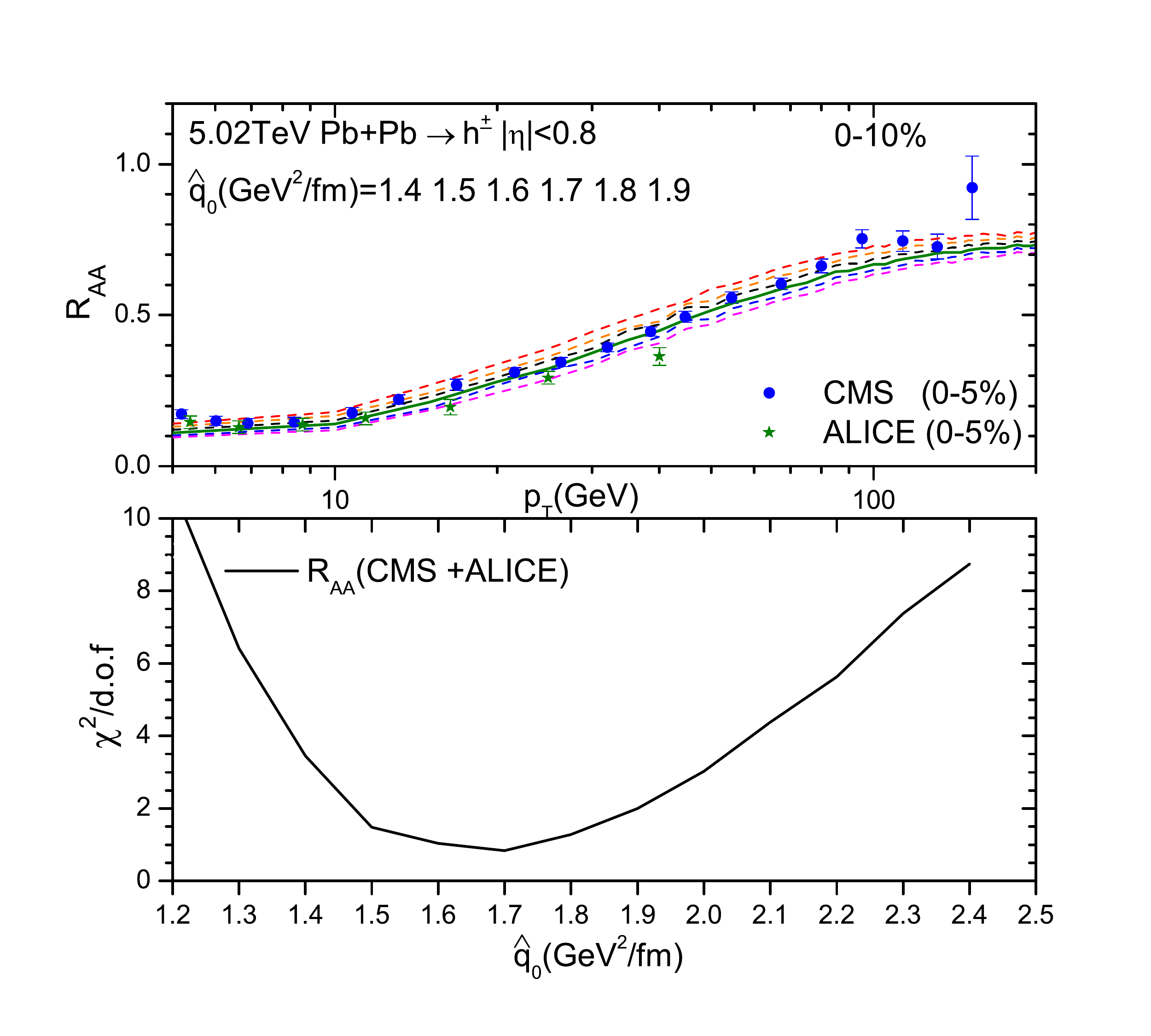}
\includegraphics[width=0.45\textwidth]{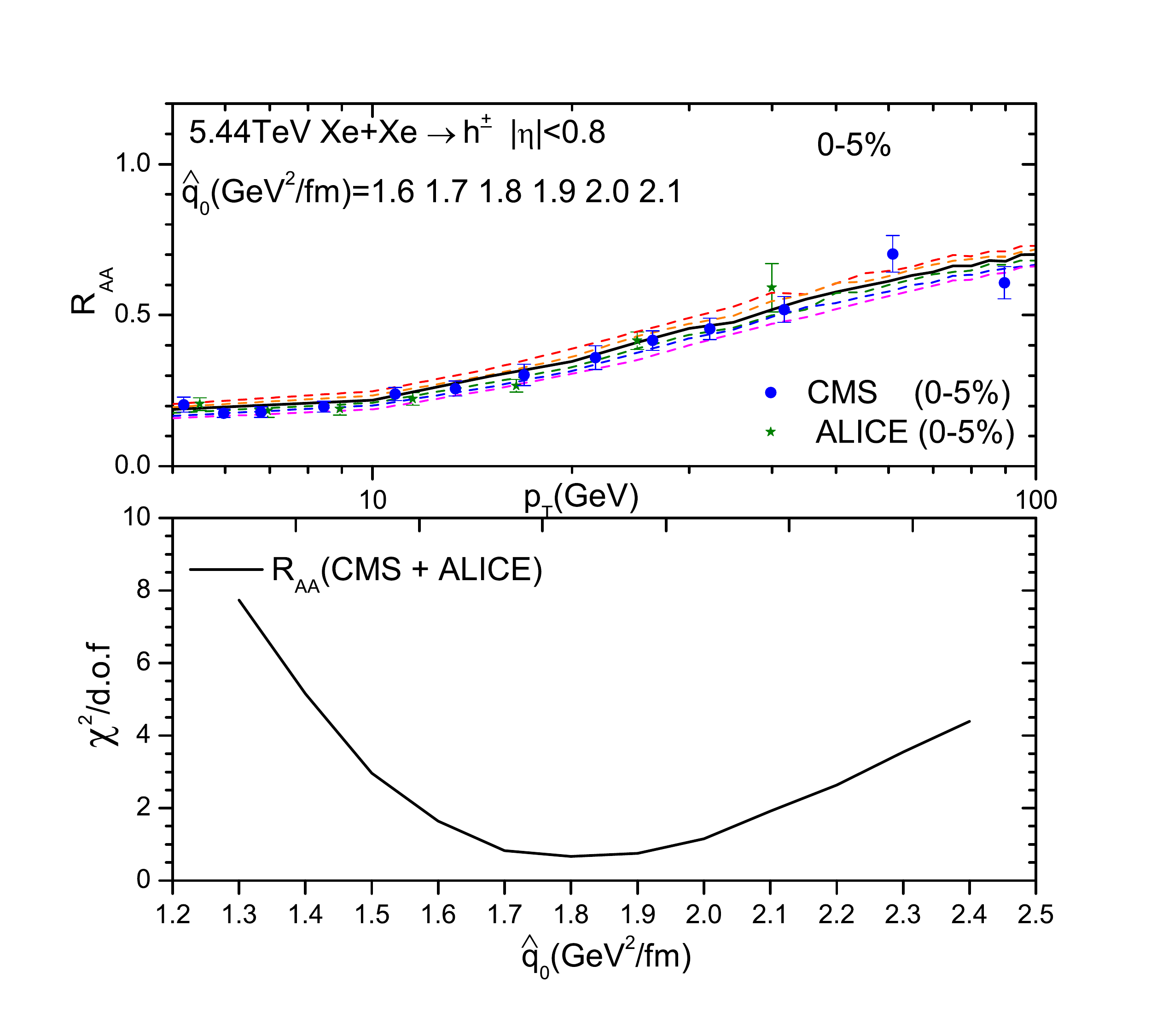}
\end{center}
\caption[*]{The single hadron suppression factors in central Pb+Pb collisions at $\sqrt{s_{\rm{NN}}}=5.02$~TeV and in central Xe+Xe collisions at $\sqrt{s_{\rm{NN}}}=5.44$~TeV compared with CMS \cite{Acharya:2018qsh,Acharya:2018eaq} and ALICE \cite{Khachatryan:2016odn} data.}
\label{fig:lhc502-544-RAA0-10}
\end{figure}

\begin{figure}[tbh]
\begin{center}
\includegraphics[width=0.45\textwidth]{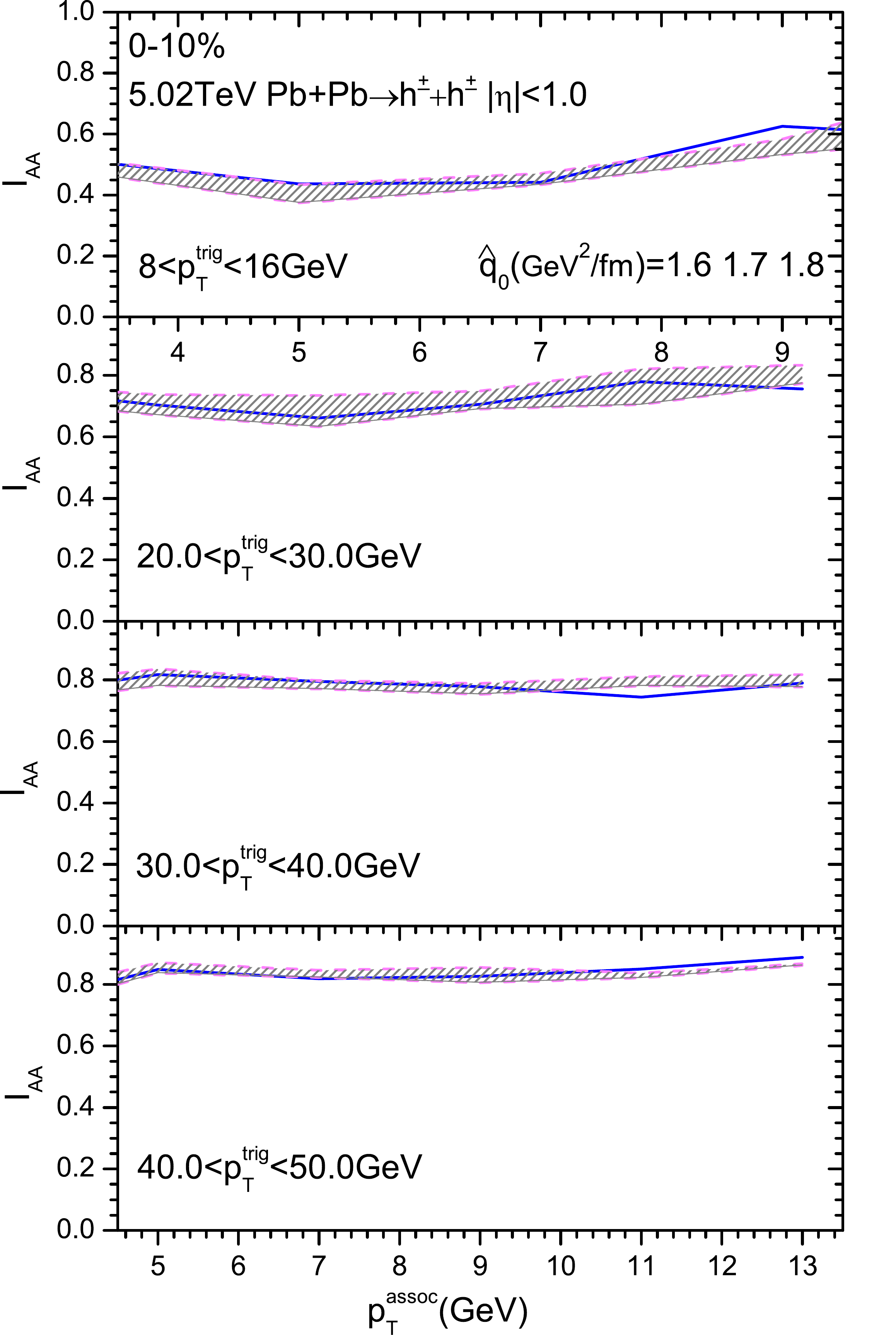}
\includegraphics[width=0.45\textwidth]{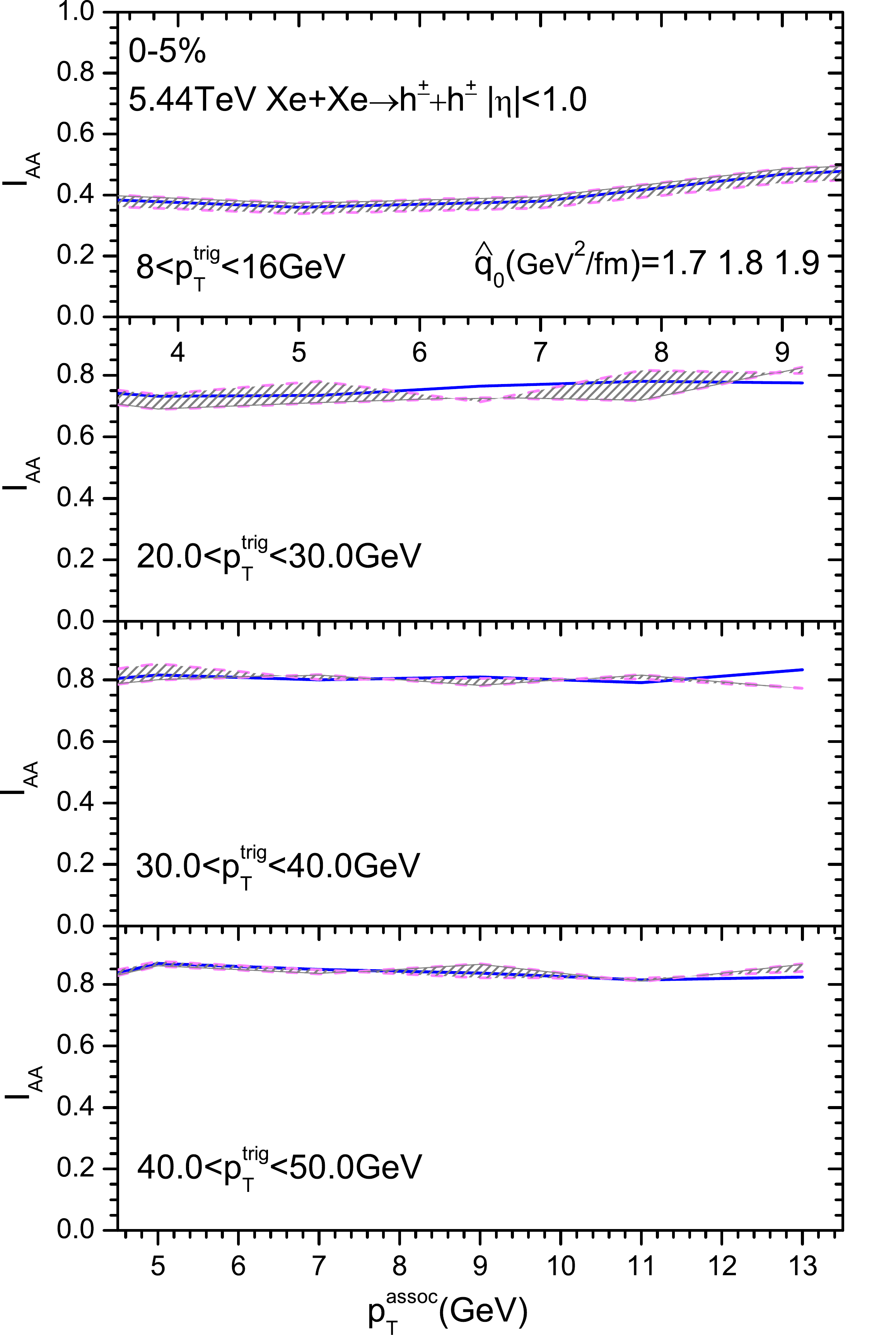}
\end{center}
	\caption[*]{The predictions of dihadron suppression factors in central Pb+Pb collisions at $\sqrt{s_{\rm{NN}}}=5.02$~TeV and in central Xe+Xe collisions at $\sqrt{s_{\rm{NN}}}=5.44$~TeV.}
\label{fig:lhc502-544-IAA0-10}
\end{figure}

\end{widetext}

\subsection{Pb+Pb collisions at $\sqrt{s_{\rm{NN}}}=5.02$~TeV and Xe+Xe collisions at $\sqrt{s_{\rm{NN}}}=5.44$~TeV at the LHC}

Recently, ALICE \cite{Khachatryan:2016odn} and CMS \cite{Acharya:2018qsh} Collaborations have published their measurements on the nuclear modification factor $R_{AA}$ for single hadron productions in Pb+Pb collisions at $\sqrt{s_{\rm{NN}}}=5.02$~TeV and Xe+Xe collisions at $\sqrt{s_{\rm{NN}}}=5.44$~TeV.
These new results provide a good opportunity for studying the collision energy and system size dependences of jet quenching in relativistic heavy-ion collisions.
Since no experimental data on dihadron nuclear modification factor $I_{AA}$ are available for these collisions, we will extract the $\hat{q}_0$ values only using the available $R_{AA}$ data.
Given that our model can provide a consistent description of both single hadron and dihadron nuclear modifications in Au+Au collisions at $\sqrt{s_{\rm{NN}}}=0.2$~TeV and Pb+Pb collisions at $\sqrt{s_{\rm{NN}}}=2.76$~TeV, we then use the extracted $\hat{q}_0$ values to predict dihadron nuclear modification factor $I_{AA}$ in Pb+Pb collisions at $\sqrt{s_{\rm{NN}}}=5.02$~TeV and Xe+Xe collisions at $\sqrt{s_{\rm{NN}}}=5.44$~TeV.

Our numerical results are shown in Fig. \ref{fig:lhc502-544-RAA0-10} and Fig. \ref{fig:lhc502-544-IAA0-10}, in which the left panels show the result for Pb+Pb collisions at $\sqrt{s_{\rm{NN}}}=5.02$~TeV and the right for Xe+Xe collisions at $\sqrt{s_{\rm{NN}}}=5.44$~TeV.
Fig. \ref{fig:lhc502-544-RAA0-10} shows the nuclear modification factor $R_{AA}$ (in the upper pannels) together with the $\chi^2$ analysis (in the lower pannels).
Again, the solid lines are the results using the best fit $\hat{q}_0$ values.
From our $\chi^2$ analysis, we obtain: $\hat{q}_0 \approx 1.7 $~GeV$^2/$fm at $T_0 = 516$~MeV ($\hat{q}/T^3 \approx 2.5 $) for central Pb+Pb collisions at  $\sqrt{s_{\rm{NN}}}=5.02$~TeV and $\hat{q}_0 \approx 1.8$~GeV$^2$/fm at $T_0=469$~MeV ($\hat{q}/T^3 \approx 3.5 $) for central Xe+Xe collisions at $\sqrt{s_{\rm{NN}}}=5.44$~TeV.
Using the extracted $\hat{q}_0$ values from fitting $R_{AA}$ data, we present in Fig. \ref{fig:lhc502-544-IAA0-10} our predictions for dihadron nuclear modification factor $I_{AA}$ in central Pb+Pb collisions at $\sqrt{s_{\rm{NN}}}=5.02$~TeV (left) and central Xe+Xe collisions at $\sqrt{s_{\rm{NN}}}=5.44$~TeV (right).
Different panels are the results with different transverse momenta for trigger hadrons.
In each plot, the solid lines in the middle are the results using the best $\hat{q}_0$ values, while the other two lines represent the theoretical which we take $\pm 0.1$~GeV$^2$/fm around the best $\hat{q}_0$ values fitted from $R_{AA}$ data.
One interesting observation is that as the values of $I_{AA}$ also increase as one increases the trigger hadron transverse momentum.
This can be understood since dihadrons with larger transverse momenta are more likely produced by tangential emissions, and thus have smaller nuclear modification effect.

\begin{figure}[tbh]
\begin{center}
\includegraphics[width=0.45\textwidth]{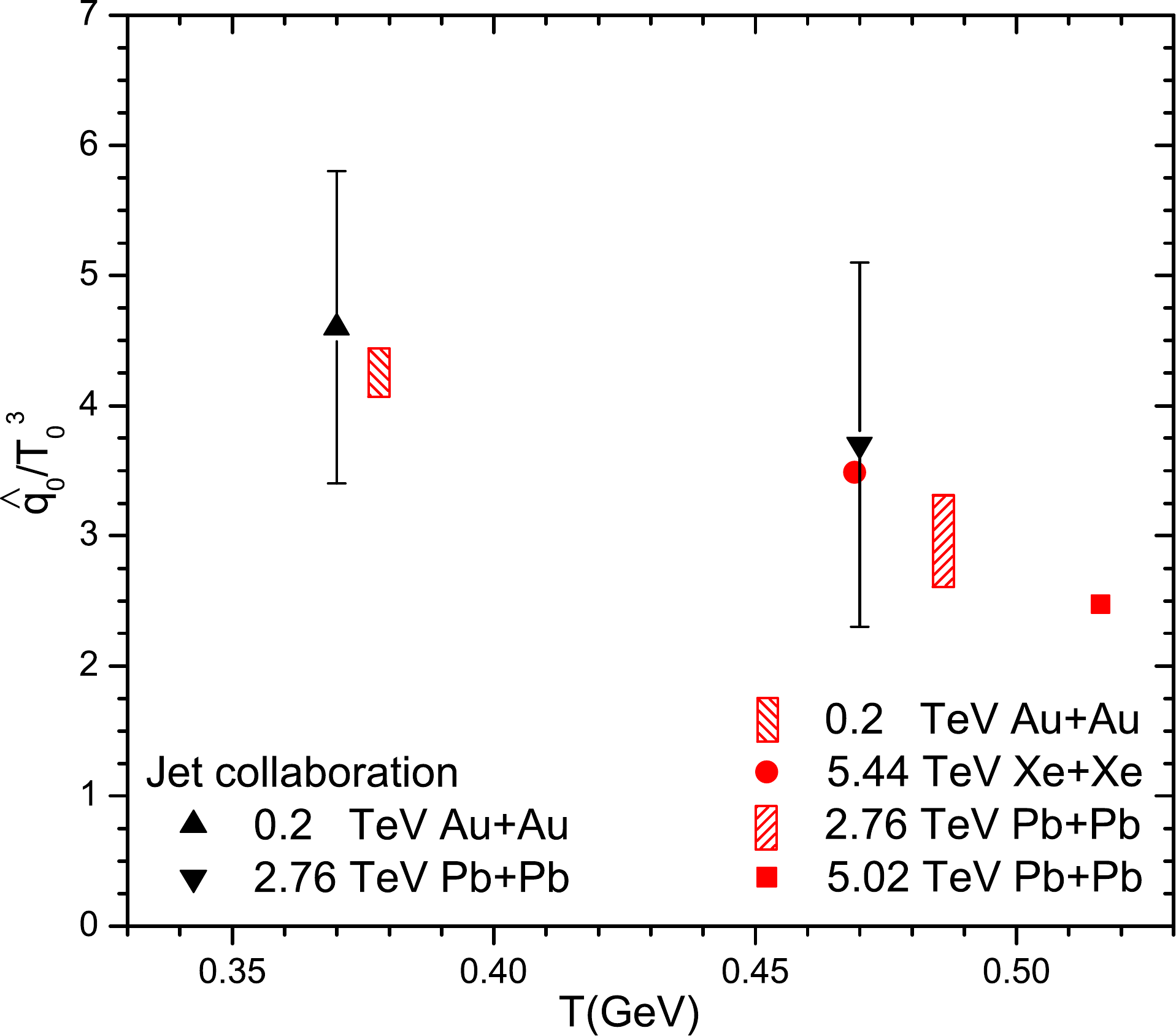}
\end{center}
	\caption[*]{The scaled jet quenching parameter $\hat{q}/T^3$ as a function of $T$ extracted via single hadron and dihadron suppression data at RHIC and the LHC. The boxes are Jet Collaboration results.}
\label{fig:JETCOLL}
\end{figure}

\subsection{$\hat{q}$ from single hadron and dihadron nuclear suppressions at RHIC and the LHC}

In previous subsections, we have quantitatively extracted the jet quenching parameter $\hat{q}$ by performing a detailed $\chi^2$ analysis on the comparison of our jet energy loss model calculations to single hadron and dihadron nuclear modification data at RHIC and the LHC. Here we summarize the main results for the extracted $\hat{q}$ values, in terms of the scaled jet quenching parameter $\hat{q}/T^3$:
\begin{eqnarray}
	&& \frac{\hat{q}}{T^3} = 4.1 \sim 4.4,\ \ T = 378~{\rm MeV};
	\nonumber\\
	&& \frac{\hat{q}}{T^3} \approx 3.5,\ \ ~~~~~~~~T = 469~{\rm MeV};
	\nonumber\\
	&& \frac{\hat{q}}{T^3} = 2.6 \sim 3.3,\ \ T = 486~{\rm MeV};
	\nonumber\\
	&& \frac{\hat{q}}{T^3} \approx 2.5,\ \ ~~~~~~~~ T = 516~{\rm MeV}.
\end{eqnarray}
One can see that the scaled jet quenching parameter $\hat{q}/T^3$ has some temperature dependence: it decreases as one increase the temperature, which may be understood as decreasing jet-medium interaction strength at higher temperature regimes.
For better visualization, we also plot the above values in Fig. \ref{fig:JETCOLL}, where the results from JET Collaboration on $\hat{q}/T^3$ for Au+Au collisions at $\sqrt{s_{\rm{NN}}}=0.2$~TeV and Pb+Pb collisions at $\sqrt{s_{\rm{NN}}}=2.76$~TeV are also shown.
We can see that our extracted values for the scaled jet quenching parameter $\hat{q}/T^3$ are consistent with the JET Collaboration results.

\section{SUMMARY}

In this work, we have studied the nuclear suppressions of single hadron and dihadron productions at high transverse momentum regimes in high-energy heavy-ion collisions at RHIC and the LHC.
We compute the cross section of single hadron and dihadron productions in relativistic nuclear collisions based on the NLO perturbative QCD framework.
For hadron production in heavy-ion collisions, we include both initial-state cold nuclear matter effect and final-state hot nuclear matter effect.
The effect of jet energy loss in hot QGP medium is taken into account using medium-modified fragmentation functions, which are calculated based on the higher-twist formalism.
The numerical results from our jet energy loss model calculations show consistent descriptions of the nuclear modifications of single hadron and dihadron productions in central and non-central nucleus-nucleus collisions at RHIC and the LHC.

We have further performed a detailed $\chi^2$ analysis by comparing our jet energy loss model calculations with the experimental data on single hadron and dihadron nuclear modifications at RHIC and the LHC.
From the global $\chi^2$ analysis, we have quantitatively extracted the values of $\hat q_0$ for different collision systems and collision energies.
For Au+Au collisions at $\sqrt{s_{\rm{NN}}}=0.2$~TeV at RHIC, we obtain $\hat{q}_0 =1.1 \sim 1.2$ GeV$^2/$fm at $T_0 =$ 378 MeV (i.e., ${\hat{q}}/{T^3} = 4.1 \sim 4.4$).
For Pb+Pb collisions at $\sqrt{s_{\rm{NN}}}=2.76$~TeV at the LHC, we obtain $\hat{q}_0 = 1.5 \sim 1.9$ GeV$^2$/fm at $T_0 =$ 486 MeV (i.e., ${\hat{q}}/{T^3} = 2.6 \sim 3.3$).
These results are consistent with the previous JET Collaboration results.
As for Pb+Pb collisions at $\sqrt{s_{\rm{NN}}} = 5.02$~TeV and Xe+Xe collisions at $\sqrt{s_{\rm{NN}}} = 5.44$~TeV, we have used single hadron $R_{AA}$ data to extract the $\hat{q}$ values.
These extracted values are then used to predict the nuclear modification effects in dihadron productions in these collisions.
Our work provides an important contribution to our quantitative extraction of the temperature dependence of jet quenching parameter by using multiple jet quenching observables from different collision systems and energies, and is helpful to achieve a consistent understanding of jet quenching in relativistic heavy-ion collisions.

\section*{ACKNOWLEDGMENTS}

This work is supported in part by Natural Science Foundation of China (NSFC) under grant Nos. 11435004, 11775095, 11890711 and 11375072.


\begin{thebibliography}{99}

\bibitem{Gyulassy:1990ye}
  M.~Gyulassy and M.~Plumer,
  Phys.\ Lett.\ B {\bf 243}, 432 (1990).
  doi:10.1016/0370-2693(90)91409-5


\bibitem{Wang:1991xy}
  X.~N.~Wang and M.~Gyulassy,
  Phys.\ Rev.\ Lett.\  {\bf 68}, 1480 (1992).
  doi:10.1103/PhysRevLett.68.1480

\bibitem{Qin:2015srf}
G.~Y.~Qin and X.~N.~Wang,
    Int.\ J.\ Mod.\ Phys.\ E {\bf 24}, no. 11, 1530014 (2015).
      doi:10.1142/S0218301315300143

\bibitem{Bass:2008rv}
  S.~A.~Bass, C.~Gale, A.~Majumder, C.~Nonaka, G.~Y.~Qin, T.~Renk and J.~Ruppert,
  Phys.\ Rev.\ C {\bf 79}, 024901 (2009)
  doi:10.1103/PhysRevC.79.024901
  [arXiv:0808.0908 [nucl-th]].


\bibitem{Armesto:2009zi}
  N.~Armesto, M.~Cacciari, T.~Hirano, J.~L.~Nagle and C.~A.~Salgado,
  J.\ Phys.\ G {\bf 37}, 025104 (2010)
  doi:10.1088/0954-3899/37/2/025104
  [arXiv:0907.0667 [hep-ph]].


\bibitem{Burke:2013yra}
  K.~M.~Burke {\it et al.} [JET Collaboration],
  Phys.\ Rev.\ C {\bf 90}, no. 1, 014909 (2014)
  doi:10.1103/PhysRevC.90.014909
  [arXiv:1312.5003 [nucl-th]].


\bibitem{Cao:2017hhk}
  S.~Cao, T.~Luo, G.~Y.~Qin and X.~N.~Wang,
  Phys.\ Lett.\ B {\bf 777}, 255 (2018)
  doi:10.1016/j.physletb.2017.12.023
  [arXiv:1703.00822 [nucl-th]].


\bibitem{Zhang:2007ja}
  H.~Zhang, J.~F.~Owens, E.~Wang and X.~N.~Wang,
  Phys.\ Rev.\ Lett.\  {\bf 98}, 212301 (2007)
  doi:10.1103/PhysRevLett.98.212301
  [nucl-th/0701045].


\bibitem{Majumder:2004pt}
  A.~Majumder, E.~Wang and X.~N.~Wang,
  Phys.\ Rev.\ Lett.\  {\bf 99}, 152301 (2007)
  doi:10.1103/PhysRevLett.99.152301
  [nucl-th/0412061].


\bibitem{Qin:2009bk}
  G.~Y.~Qin, J.~Ruppert, C.~Gale, S.~Jeon and G.~D.~Moore,
  Phys.\ Rev.\ C {\bf 80}, 054909 (2009)
  doi:10.1103/PhysRevC.80.054909
  [arXiv:0906.3280 [hep-ph]].


\bibitem{Renk:2008xq}
  T.~Renk,
  Phys.\ Rev.\ C {\bf 78}, 034904 (2008)
  doi:10.1103/PhysRevC.78.034904
  [arXiv:0803.0218 [hep-ph]].


\bibitem{Chen:2016vem}
  L.~Chen, G.~Y.~Qin, S.~Y.~Wei, B.~W.~Xiao and H.~Z.~Zhang,
  Phys.\ Lett.\ B {\bf 773}, 672 (2017)
  doi:10.1016/j.physletb.2017.09.031
  [arXiv:1607.01932 [hep-ph]].


\bibitem{Chen:2017zte}
  W.~Chen, S.~Cao, T.~Luo, L.~G.~Pang and X.~N.~Wang,
  Phys.\ Lett.\ B {\bf 777}, 86 (2018)
  doi:10.1016/j.physletb.2017.12.015
  [arXiv:1704.03648 [nucl-th]].


\bibitem{Qin:2010mn}
  G.~Y.~Qin and B.~Muller,
  Phys.\ Rev.\ Lett.\  {\bf 106}, 162302 (2011)
  Erratum: [Phys.\ Rev.\ Lett.\  {\bf 108}, 189904 (2012)]
  doi:10.1103/PhysRevLett.108.189904, 10.1103/PhysRevLett.106.162302
  [arXiv:1012.5280 [hep-ph]].


\bibitem{CasalderreySolana:2010eh}
  J.~Casalderrey-Solana, J.~G.~Milhano and U.~A.~Wiedemann,
  J.\ Phys.\ G {\bf 38}, 035006 (2011)
  doi:10.1088/0954-3899/38/3/035006
  [arXiv:1012.0745 [hep-ph]].


\bibitem{He:2011pd}
  Y.~He, I.~Vitev and B.~W.~Zhang,
  Phys.\ Lett.\ B {\bf 713}, 224 (2012)
  doi:10.1016/j.physletb.2012.05.054
  [arXiv:1105.2566 [hep-ph]].


\bibitem{Young:2011qx}
  C.~Young, B.~Schenke, S.~Jeon and C.~Gale,
  Phys.\ Rev.\ C {\bf 84}, 024907 (2011)
  doi:10.1103/PhysRevC.84.024907
  [arXiv:1103.5769 [nucl-th]].


\bibitem{Zapp:2012ak}
  K.~C.~Zapp, F.~Krauss and U.~A.~Wiedemann,
  JHEP {\bf 1303}, 080 (2013)
  doi:10.1007/JHEP03(2013)080
  [arXiv:1212.1599 [hep-ph]].


\bibitem{Wang:2013cia}
  X.~N.~Wang and Y.~Zhu,
  Phys.\ Rev.\ Lett.\  {\bf 111}, no. 6, 062301 (2013)
  doi:10.1103/PhysRevLett.111.062301
  [arXiv:1302.5874 [hep-ph]].


\bibitem{Chang:2016gjp}
  N.~B.~Chang and G.~Y.~Qin,
  Phys.\ Rev.\ C {\bf 94}, no. 2, 024902 (2016)
  doi:10.1103/PhysRevC.94.024902
  [arXiv:1603.01920 [hep-ph]].


\bibitem{Tachibana:2017syd}
  Y.~Tachibana, N.~B.~Chang and G.~Y.~Qin,
  Phys.\ Rev.\ C {\bf 95}, no. 4, 044909 (2017)
  doi:10.1103/PhysRevC.95.044909
  [arXiv:1701.07951 [nucl-th]].


\bibitem{Baier:1996sk}
  R.~Baier, Y.~L.~Dokshitzer, A.~H.~Mueller, S.~Peigne and D.~Schiff,
  Nucl.\ Phys.\ B {\bf 484}, 265 (1997)
  doi:10.1016/S0550-3213(96)00581-0
  [hep-ph/9608322].


\bibitem{Baier:1996kr}
  R.~Baier, Y.~L.~Dokshitzer, A.~H.~Mueller, S.~Peigne and D.~Schiff,
  Nucl.\ Phys.\ B {\bf 483}, 291 (1997)
  doi:10.1016/S0550-3213(96)00553-6
  [hep-ph/9607355].


\bibitem{Baier:1998kq}
  R.~Baier, Y.~L.~Dokshitzer, A.~H.~Mueller and D.~Schiff,
  Nucl.\ Phys.\ B {\bf 531}, 403 (1998)
  doi:10.1016/S0550-3213(98)00546-X
  [hep-ph/9804212].


\bibitem{Guo:2000nz}
  X.~f.~Guo and X.~N.~Wang,
  Phys.\ Rev.\ Lett.\  {\bf 85}, 3591 (2000)
  doi:10.1103/PhysRevLett.85.3591
  [hep-ph/0005044].


\bibitem{Wang:2001ifa}
  X.~N.~Wang and X.~f.~Guo,
  Nucl.\ Phys.\ A {\bf 696}, 788 (2001)
  doi:10.1016/S0375-9474(01)01130-7
  [hep-ph/0102230].


\bibitem{Majumder:2009ge}
  A.~Majumder,
  Phys.\ Rev.\ D {\bf 85}, 014023 (2012)
  doi:10.1103/PhysRevD.85.014023
  [arXiv:0912.2987 [nucl-th]].


\bibitem{Andres:2016iys}
  C.~Andrés, N.~Armesto, M.~Luzum, C.~A.~Salgado and P.~Zurita,
  Eur.\ Phys.\ J.\ C {\bf 76}, no. 9, 475 (2016)
  doi:10.1140/epjc/s10052-016-4320-5
  [arXiv:1606.04837 [hep-ph]].


\bibitem{Andres:2016uio}
  C.~Andrés, N.~Amesto, M.~Luzum, C.~A.~Salgado and P.~Zurita,
  Nucl.\ Part.\ Phys.\ Proc.\  {\bf 289-290}, 105 (2017)
  doi:10.1016/j.nuclphysbps.2017.05.020
  [arXiv:1612.06781 [nucl-th]].


\bibitem{Andres:2017awo}
  C.~Andres, N.~Armesto, H.~Niemi, R.~Paatelainen, C.~A.~Salgado and P.~Zurita,
  Nucl.\ Phys.\ A {\bf 967}, 492 (2017)
  doi:10.1016/j.nuclphysa.2017.05.115
  [arXiv:1705.01493 [nucl-th]].


\bibitem{Adare:2008qa}
  A.~Adare {\it et al.} [PHENIX Collaboration],
  Phys.\ Rev.\ Lett.\  {\bf 101}, 232301 (2008)
  doi:10.1103/PhysRevLett.101.232301
  [arXiv:0801.4020 [nucl-ex]].


\bibitem{Adare:2012wg}
  A.~Adare {\it et al.} [PHENIX Collaboration],
  Phys.\ Rev.\ C {\bf 87}, no. 3, 034911 (2013)
  doi:10.1103/PhysRevC.87.034911
  [arXiv:1208.2254 [nucl-ex]].


\bibitem{Adams:2006yt}
  J.~Adams {\it et al.} [STAR Collaboration],
  Phys.\ Rev.\ Lett.\  {\bf 97}, 162301 (2006)
  doi:10.1103/PhysRevLett.97.162301
  [nucl-ex/0604018].


\bibitem{STAR:2016jdz}
  L.~Adamczyk {\it et al.} [STAR Collaboration],
  Phys.\ Lett.\ B {\bf 760}, 689 (2016)
  doi:10.1016/j.physletb.2016.07.046
  [arXiv:1604.01117 [nucl-ex]].


\bibitem{Abelev:2012hxa}
  B.~Abelev {\it et al.} [ALICE Collaboration],
  Phys.\ Lett.\ B {\bf 720}, 52 (2013)
  doi:10.1016/j.physletb.2013.01.051
  [arXiv:1208.2711 [hep-ex]].


\bibitem{CMS:2012aa}
  S.~Chatrchyan {\it et al.} [CMS Collaboration],
  Eur.\ Phys.\ J.\ C {\bf 72}, 1945 (2012)
  doi:10.1140/epjc/s10052-012-1945-x
  [arXiv:1202.2554 [nucl-ex]].


\bibitem{Khachatryan:2016odn}
  V.~Khachatryan {\it et al.} [CMS Collaboration],
  JHEP {\bf 1704}, 039 (2017)
  doi:10.1007/JHEP04(2017)039
  [arXiv:1611.01664 [nucl-ex]].


\bibitem{Acharya:2018qsh}
  S.~Acharya {\it et al.} [ALICE Collaboration],
  JHEP {\bf 1811}, 013 (2018)
  doi:10.1007/JHEP11(2018)013
  [arXiv:1802.09145 [nucl-ex]].


\bibitem{Acharya:2018eaq}
  S.~Acharya {\it et al.} [ALICE Collaboration],
  Phys.\ Lett.\ B {\bf 788}, 166 (2019)
  doi:10.1016/j.physletb.2018.10.052
  [arXiv:1805.04399 [nucl-ex]].


\bibitem{Adam:2016xbp}
  J.~Adam {\it et al.} [ALICE Collaboration],
  Phys.\ Lett.\ B {\bf 763}, 238 (2016)
  doi:10.1016/j.physletb.2016.10.048
  [arXiv:1608.07201 [nucl-ex]].


\bibitem{Aamodt:2011vg}
  K.~Aamodt {\it et al.} [ALICE Collaboration],
  Phys.\ Rev.\ Lett.\  {\bf 108}, 092301 (2012)
  doi:10.1103/PhysRevLett.108.092301
  [arXiv:1110.0121 [nucl-ex]].


\bibitem{Conway:2013xaa}
  R.~Conway [CMS Collaboration],
  Nucl.\ Phys.\ A {\bf 904-905}, 451c (2013).
  doi:10.1016/j.nuclphysa.2013.02.046



\bibitem{Hou:2016nqm}
  T.~J.~Hou {\it et al.},
  Phys.\ Rev.\ D {\bf 95}, no. 3, 034003 (2017)
  doi:10.1103/PhysRevD.95.034003
  [arXiv:1609.07968 [hep-ph]].

\bibitem{Kretzer:2000yf}
  S.~Kretzer,
  Phys.\ Rev.\ D {\bf 62}, 054001 (2000)
  doi:10.1103/PhysRevD.62.054001
  [hep-ph/0003177].


\bibitem{Wang:2004yv}
  X.~N.~Wang,
  Phys.\ Rev.\ C {\bf 70}, 031901 (2004)
  doi:10.1103/PhysRevC.70.031901
  [nucl-th/0405029].


\bibitem{Zhang:2009rn}
  H.~Zhang, J.~F.~Owens, E.~Wang and X.~N.~Wang,
  Phys.\ Rev.\ Lett.\  {\bf 103}, 032302 (2009)
  doi:10.1103/PhysRevLett.103.032302
  [arXiv:0902.4000 [nucl-th]].


\bibitem{Wang:2003mm}
  X.~N.~Wang,
  Phys.\ Lett.\ B {\bf 595}, 165 (2004)
  doi:10.1016/j.physletb.2004.05.021
  [nucl-th/0305010].


\bibitem{Wang:1996yf}
  X.~N.~Wang,
  Phys.\ Rept.\  {\bf 280}, 287 (1997)
  doi:10.1016/S0370-1573(96)00022-1
  [hep-ph/9605214].


\bibitem{Li:2001xa}
  S.~y.~Li and X.~N.~Wang,
  Phys.\ Lett.\ B {\bf 527}, 85 (2002)
  doi:10.1016/S0370-2693(02)01179-6
  [nucl-th/0110075].

\bibitem{Emelyanov:1999pkc}
  V.~Emel'yanov, A.~Khodinov, S.~R.~Klein and R.~Vogt,
  Phys.\ Rev.\ C {\bf 61}, 044904 (2000)
  doi:10.1103/PhysRevC.61.044904
  [hep-ph/9909427].


\bibitem{Hirano:2003pw}
  T.~Hirano and Y.~Nara,
  Phys.\ Rev.\ C {\bf 69}, 034908 (2004)
  doi:10.1103/PhysRevC.69.034908
  [nucl-th/0307015].

\bibitem{Eskola:2016oht}
  K.~J.~Eskola, P.~Paakkinen, H.~Paukkunen and C.~A.~Salgado,
  Eur.\ Phys.\ J.\ C {\bf 77}, no. 3, 163 (2017)
  doi:10.1140/epjc/s10052-017-4725-9
  [arXiv:1612.05741 [hep-ph]].



\bibitem{Wang:2009qb}
  W.~t.~Deng and X.~N.~Wang,
  Phys.\ Rev.\ C {\bf 81}, 024902 (2010)
  doi:10.1103/PhysRevC.81.024902
  [arXiv:0910.3403 [hep-ph]].


\bibitem{Wang:2001cs}
  E.~Wang and X.~N.~Wang,
  Phys.\ Rev.\ Lett.\  {\bf 87}, 142301 (2001)
  doi:10.1103/PhysRevLett.87.142301
  [nucl-th/0106043].


\bibitem{Wang:2002ri}
  E.~Wang and X.~N.~Wang,
  Phys.\ Rev.\ Lett.\  {\bf 89}, 162301 (2002)
  doi:10.1103/PhysRevLett.89.162301
  [hep-ph/0202105].


\bibitem{Chang:2014fba}
  N.~B.~Chang, W.~T.~Deng and X.~N.~Wang,
  Phys.\ Rev.\ C {\bf 89}, no. 3, 034911 (2014)
  doi:10.1103/PhysRevC.89.034911
  [arXiv:1401.5109 [nucl-th]].


\bibitem{Song:2007fn}
  H.~Song and U.~W.~Heinz,
  Phys.\ Lett.\ B {\bf 658}, 279 (2008)
  doi:10.1016/j.physletb.2007.11.019
  [arXiv:0709.0742 [nucl-th]].


\bibitem{Song:2007ux}
  H.~Song and U.~W.~Heinz,
  Phys.\ Rev.\ C {\bf 77}, 064901 (2008)
  doi:10.1103/PhysRevC.77.064901
  [arXiv:0712.3715 [nucl-th]].


\bibitem{Qiu:2011hf}
  Z.~Qiu, C.~Shen and U.~Heinz,
  Phys.\ Lett.\ B {\bf 707}, 151 (2012)
  doi:10.1016/j.physletb.2011.12.041
  [arXiv:1110.3033 [nucl-th]].


\bibitem{Qiu:2012uy}
  Z.~Qiu and U.~Heinz,
  Phys.\ Lett.\ B {\bf 717}, 261 (2012)
  doi:10.1016/j.physletb.2012.09.030
  [arXiv:1208.1200 [nucl-th]].


\bibitem{Wang:2003aw}
  X.~N.~Wang,
  Phys.\ Lett.\ B {\bf 579}, 299 (2004)
  doi:10.1016/j.physletb.2003.11.011
  [nucl-th/0307036].

\end{thebibliography}

\end{document}